# Reconciling Complexity and Simplicity in the Business Model Canvas Design Through Metamodelling and Domain-Specific Modelling

Nordine Benkeltoum
Lecturer in Management Information Systems
Centrale Lille Institut
CS 20048
59651 Villeneuve d'Ascq cedex
France
nordine.benATgmail.com

## Abstract

This article introduces a metamodel for the Business Model Canvas (BMC) using the Unified Modelling Language (UML), together with a dedicated Domain-Specific Modelling Language (DSML) tool. Although the BMC is widely adopted by both practitioners and scholars, significant challenges remain in formally modelling business models, particularly with regard to explicit specification of inter-component relationships, while preserving the simplicity that characterises the BMC. Addressing this tension between modelling rigour and practical relevance, this research adopts a Design Science Research approach to formally specify relationships among BMC components and to strengthen their theoretical grounding through an adaptation of the $V^4$ framework. The proposed metamodel consolidates BMC relationships into three core types: *supports*, *determines*, and *affects*, providing explicit semantics while remaining accessible to end users through graphical tooling. The findings highlight that formally specifying relationships significantly improves the interpretability and consistency of BMC representations. The proposed metamodel and tool offer a rigorous yet usable foundation for developing DSML-based BMC tools and for enabling systematic integration of the BMC into widely used software and enterprise modelling environments, thereby bridging business modelling and enterprise architecture practices for both academics and practitioners.

**Introduction**

The business model has become an increasingly prominent topic among leading practitioners. According to a recent international survey, four in ten CEOs believe their company may not remain viable beyond the next ten years if it continues on its current trajectory. This finding underscores the ongoing need for continuous change in the contemporary economy (PWC, 2025, p. 11). This aligns with current academic insights from the Resource-Based View (RBV). Without being supported by Valuable, Rare, Inimitable, and Non-substitutable (VRIN) resources (Barney, 1991), a business model can be easily replicated (Teece, 2010, p. 192). In turbulent environments, merely possessing strategic resources is not enough; firms must develop dynamic capabilities (Teece, 2007, p. 1319). The business model has also gained traction within the Information Systems (IS) research community (Lara Machado et al., 2024). To illustrate this trend, over the past five years (since 5 June 2020), no fewer than 16,911 articles in Springer Nature Link have mentioned the term "*business model*" in keywords.

The business model concept is considered a key concept linking strategy, processes, and information technologies (Veit et al., 2014). Nevertheless, the relationship between strategy and business model remains unclear (Kulins et al., 2016, p. 1437) although they are distinct concepts closely linked concepts (Veit et al., 2014, p. 46). Some scholars suggest that the business model is an abstraction of strategy (Seddon et al., 2004) while others argue that it acts as a bridge between strategy and operations (Al-Debei & Avison, 2010; Feller et al., 2011; Lara Machado et al., 2024). Other researchers argue that the connection between business model and operational implementation remains insufficiently understood (Lara Machado et al., 2024, p. 628). This article examines the Business Model Canvas framework (Osterwalder & Pigneur, 2010). More than one million copies of the Business Model Canvas seminal book have been sold. The business model has also attracted considerable interest among IS scholars. The Business Model Canvas (BMC) appears to be the most widely cited conceptual framework in IS literature. Despite its clear structure, the use of BMC framework still faces significant challenges. The first challenge concerns the relationships between components, which are only partially specified. In practice, a large proportion of practitioners uses the BMC without paying attention to relationships between components as stressed in previous research (Avdiji et al., 2020, p. 708). The lack of specified BMC relationships is highlighted in IS research (Curty & Fill, 2022, p. 4; Wieland & Fill, 2020, p. 47). Besides prior research emphasises that practical utility of the business model concept is contingent upon formalisation (Veit et al., 2014, p. 46). Consequently, this article addresses the following research question:

*How can relevance and rigour be reconciled when modelling the relationships between Business Model Canvas components using complex metamodelling and simple domain-specific modelling language tooling?*

This article is divided into five parts. The next section reviews the literature on the Business Model Canvas and the Resource-Based View. The following section presents the theoretical foundations of the Business Model Canvas and business model formalisations. The subsequent section describes Design Science Research methodology and the metamodelling rationale. This is followed by an evaluation of research artefacts. Finally, the paper concludes with a discussion of the main contributions.

## 1. The Business Model Canvas and the Resource-Based View
### 1.1. From a Static to a Dynamic Business Model Canvas perspective

The business model (BM) lacks a universally accepted definition (DaSilva & Trkman, 2014; Fritscher & Pigneur, 2015; Lara Machado et al., 2024). It also lacks a solid theoretical foundation (Al-Debei & Avison, 2010, p. 364). Several scholars consider it an important concept but still vague or obscure (Al-Debei & Avison, 2010, p. 359; Hedman & Kalling, 2003, p. 49). As a result, BMs have been defined in various ways such as: a framework, a description, a representation, a tool, a method, or even as an architecture[1] (Zott et al., 2011, p. 2023). In summary, a BM is a representation (Al-Debei & Avison, 2010, p. 372) of the way a company creates, delivers, and captures value (Osterwalder & Pigneur, 2010, p. 14). This is the definition adopted by this research.

In Information Systems (IS) research, BM is often defined as an architecture that connects strategy and operations (Al-Debei & Avison, 2010, p. 370; Feller et al., 2011, p. 360). One key distinction between strategy and business model, is the fact that strategy tends to be relatively stable over time whereas business models are inherently dynamic (Lara Machado et al., 2024, p. 609). IS contributes to BMs knowledge in three ways: through strategic modelling, by considering strategy as a design activity, and by using computer-aided design (Osterwalder & Pigneur, 2013). Moreover, the IS community provides various artefacts (models, methods, and software) to assist business model design (Lara Machado et al., 2024, p. 609). In IS, the Business Model Canvas (BMC) framework appears to be the most widely used theoretical lens to explore BM-related questions. It is considered a *quasi-standard* (Avdiji et al., 2020; Lara Machado et al., 2024). The BMC serves as a language to describe, visualise, assess, and modify business models (Osterwalder & Pigneur, 2010). For instance, BMC has been used as a holistic framework to study strategic positioning and operations in the public sector (Feller et al., 2011, p. 371).

Prior studies have mainly adopted a static perspective on the BMC. BMC is composed of nine concepts, namely: *Key Resources* (*KR*), *Key Activities* (*KA*), *Key Partnerships* (*KP*), *Customer Segments* (*CS*), *Value Proposition* (*VP*), *Channels* (*CH*), *Customer Relationships* (*CR*), *Revenue*

---

*Streams* (*R\$*), and *Cost Structure* (*C\$*). Table 1 offers a summary of each building block (Fritscher & Pigneur, 2011, 2015).

| Concepts | Abbreviation | Definition |
|---|---|---|
| Key Resources | KR | The Key Resources (KR) refer to the elements (tangible, intangible or financial) on which the company builds its value proposition. |
| Key Activities | KA | The Key Activities (KA) enable the transformation of resources into final products or services. |
| Key Partnerships | KP | The Key Partnerships (KP) represent relationships with external actors without which the company cannot fulfil its value proposition. |
| Customer Segments | CS | The Customer Segments (CS) describe clients divided into groups with the aim of identifying their needs, desires, and goals. |
| Value Proposition | VP | The Value Proposition (VP) describes the problem being solved by the company and why its product/service is superior to those of its competitors. |
| Channels | CH | The distribution Channels (CH) refer to how the clients want to be contacted and by whom. |
| Customer Relationship | CR | The Customer Relationship (CR) describes the way that the relationship is established and maintained with each customer segment. |
| Revenue Streams | R\$ | The Revenue Streams (R\$) refer to the value the clients agree to pay to acquire products and services of the company. |
| Cost Structure | C\$ | The Cost Structure (C\$) describes cost elements that allow the company to realise its value proposition. |

*Table 1 : Business Canvas Model concepts*

The nine building blocks are interconnected. The concept of *pattern* is used to describe this relationships (Osterwalder & Pigneur, 2010, p. 55; Osterwalder et al., 2020, p. 130). A pattern is an accepted model to imitate. It is used in many disciplines such as music, psychology, or even art (Coad, 1992). Five BMC patterns are suggested[2] (Table 2).

| Pattern Name | Business model consists in… |
|---|---|
| Unbundling Business Models | Managing relationship, product, and/or infrastructure-centred businesses. |
| The Long Tail | Selling a large number of different products but in limited quantities of each. |
| Multi-Sided Platforms | Addressing markets that are interconnected. |
| Free as Business Model | Interacting with primarily one customer segment that benefits a free product. |
| Open Business Models | Using external partners to leverage internal ideas or to explore new ideas. |

*Table 2 : BMC pattens*

However, the patterns do not clearly describe the causal relationships between the BMC components. They represent *visual* rather than *engineering patterns*. The concept of *pattern* is widely used within the IS community. Patterns are logical sequences that can be used to solve similar problems or to design a solution based on a commonly accepted model (Coad, 1992, p. 158; Johnson, 1997, p. 41).

---

[2] Other visual patterns are suggested in Osterwalder, A., Pigneur, Y., Smith, A., & Etiemble, F. (2020). *The Invincible Company: How to Constantly Reinvent Your Organization with Inspiration From the World's Best Business Models*. Wiley. https://books.google.fr/books?id=CQQLtQEACAAJ .

The BMC has also been studied from a design science perspective (Fritscher & Pigneur, 2010, 2011, 2015). This line of research proposes a business model ontology based on the BMC concepts. The BMC components are grouped into four perspectives, namely: the activity perspective (KP, KR, KA), the product/service perspective (VP), the customer perspective (CS, CH, CR) and the financial perspective (C$, R$). Sixteen relationships between components have been explicitly labelled. Table 3 summarises the relationships identified in previous research (Fritscher & Pigneur, 2010).

| Components | Relationship |
|---|---|
| KP → KA | performs |
| KA → KR | builds on |
| KP → KR | provides |
| KA → VP | enables |
| KP → VP | contributes |
| KR → VP | is required by |
| CR → VP | is supported by |
| CR → CS | addresses |
| CH → CS | reaches |
| CH → R$ | generates |
| CS → R$ | generates |
| VP → CS | targets |
| CR → CH | established through |

*Table 3: relationships specified in previous research*

Although a business model is often defined as a system composed of components and linkages between them (Afuah & Tucci, 2003, p. 4; Zott et al., 2011) or as articulation between components (Teece, 2007, p. 1329), representing business model dynamics remains a challenge (DaSilva & Trkman, 2014). This article considers BMC as the foundation of a Domain-Specific Modelling Language (Dietz & Juhrisch, 2012), following previous research (Fill, 2020; Wieland & Fill, 2020). While the relationships between BMC components remain unclear, this article seeks to clarify these relationships in a structured and theoretically grounded manner. The contribution of this article is to improve the formal nature of the BMC particularly since it is extensively used in IS research.

## 1.2. Resource-Based View as a Theoretical Foundation for the Business Model Canvas

The two most influential paradigms in strategy theory are the Industrial Organisation (IO) and the Resource-Based View (RBV) perspectives. The main opposition between IO and RBV concerns the explanation of the sources of competitive advantage (Hedman & Kalling, 2003). While the IO perspective explains performance through external factors (e.g. customer bargaining power, entry barriers, supplier pressure), the RBV focuses on internal factors (i.e. the resources) while also considering external resources. RBV scholars consider that market structure is the result of innovation and learning while IO theorists attribute causality in the opposite direction (Teece, 2007, p. 1325). Nevertheless, some scholars consider the RBV and the IO rather complementary (Kraaijenbrink et al., 2010). Resources are defined as a set of assets and capabilities that enable the firm to identify and respond to environmental opportunities and threats (Wade & Hulland, 2004, p. 3). The RBV regards the company as a bundle of resources that enables its strategic capabilities (Barney, 1991). Hence, business enterprises consist of portfolios of assets and competencies (Teece, 2007, p. 1319; Wade & Hulland, 2004, p. 3). The RBV is more applicable to relatively stable industries than to turbulent environments, where technologies and markets are rapidly evolving (Kraaijenbrink et al., 2010, p. 353). In turbulent environments neither resource ownership nor mastery of critical success factors is sufficient. Rather, leading companies require dynamic capabilities (Teece, 2007, pp. 1319-1320). The resources can be owned, controlled, or mobilised by the company (through partnerships). However, the distinction between resource types need to be clarified (Kraaijenbrink et al., 2010, p. 359), resources are usually classified into categories such as tangible, intangible, financial, human-based, organisational, technological, relational, or even reputational. Resources may also be categorised in terms of competitiveness: some provide competitive advantages (*ex ante*), while others resources help sustain them over time (*ex post*) (Wade & Hulland, 2004, p. 9).

The Valuable, Rare, Inimitable, and Non-substitutable (VRIN) Framework enables resource assessment to identify sources of competitive advantage (Barney & Wright, 1998, p. 32). A resource is considered *valuable* when it contributes to improvements in efficiency and effectiveness. A resource is considered *rare* when it is not available to a large number of companies. Resource *imitability* means that it is difficult to copy. Resource *substitutability* refers to available resource that is strategically equivalent (Wade & Hulland, 2004, pp. 9-11). According to the RBV, achieving Sustained Competitive Advantage (SCA) requires acquiring and controlling VRIN resources and capabilities, as well as having the organisation to absorb and apply them effectively (Kraaijenbrink et al., 2010, p. 350).

Even if scholars call for a clearer distinction between what is (not) a resource (Kraaijenbrink et al., 2010, p. 359),  it is evident that not all resources hold the same level of strategic importance. Some resources serve as sources of SCA, while others may offer only temporary advantage or may not significantly enhance competitive positioning. Such resource can often be sold or replaced by outsourced services. Some resources can be easily bought and sold while others are less easily bought and sold (Wade & Hulland, 2004, p. 10). Furthermore, resources combination is seen as a way to create distinctive and rare assets. In this context, resources should not be considered separately but rather from a holistic perspective, through resource configuration (Al-Debei & Avison, 2010, p. 367). In turbulent environment, dynamic capabilities refer to select and develop technologies and business models through the combination and management of unique assets (Teece, 2007, p. 1325).

## 2. Building a Business Model Canvas Metamodel: Theoretical Foundations
### 2.1. Business Model Formalisations

More than 17 business model representations have been identified in previous research (Szopinski et al., 2022); this research focuses on the most popular ones in IS research. A first representation is the e3value Model (Gordijn et al., 2000) focused on business transactions. The e3value model focuses on the type of objects that need to be transferred to enable transactions between exchange parties. Customers and suppliers are considered as equal from this perspective. The main difference between the e3value and the Business Model Canvas (BMC) is that e3value focuses on value transactions, whereas the BMC adopts a holistic perspective on the business model (Razo-Zapata et al., 2015).

The Business Motivation Model (BMM) represents the motivations of a company (Bernaert et al., 2016). BMM consists of *Ends* (vision and desired results), *Means* (course of action and directives), and *Influencers* (external and internal) (Object Management Group, 2015). Some of the components of the BMM are integrated into ArchiMate language (e.g. Course of Action, Goal) while others are not implemented (e.g. vision, mission) (The Open Group, 2023, p. 176). BMM is focused on underlying motivations of a company, whereas BMC adopts a more value creation-centred perspective.

Another stream of research proposes mappings between the BMC and ArchiMate. Early contributions are based on ArchiMate version 1 (Fritscher & Pigneur, 2011, 2015; Meertens et al., 2012), while more recent studies offer a mapping relying on ArchiMate version 3.1 (Iacob et al., 2014; Koyama et al., 2023; Walters, 2020). However, the literature remains fragmented, and the proposed mappings are often contradictory. For instance, *Key Activity* (*KA*) is mapped to *Business Process* and *Actor/Role* concepts by (Fritscher & Pigneur, 2011), to *Capability* by (Iacob et al., 2014), to *Business Process* by (Koyama et al., 2023), and to *Business Function* by (Walters, 2020). These inconsistencies cannot be attributed to ArchiMate, which is a rigorously specified architecture language, but rather to the BMC itself, which was originally specified as a visual and strategic management tool (Avdiji et al., 2020). Therefore, providing a rigorous mapping requires a clear BMC specification. Accordingly, this research is focused on the BMC specification itself, rather than on its integration into existing enterprise architecture modelling frameworks such as ArchiMate.

The last stream of research seeks to represent and formalise the BMC. Most BMC tools offer primarily visual representations (Wieland & Fill, 2020, p. 47) such as Visual Inquiry Tools (VIT), which support strategic decision-making (Avdiji et al., 2020). These studies aim to overcome the limitations of visual modelling tools. The Business Transaction Model (BTM) provides a semi-formal representation of business models by defining explicit relationships between BMC building blocks (Fill, 2020). The identified relationships are the following: *delivers value*, *causes*, *is part of*, *is offered*, *is used*, and *pays* (Wieland & Fill, 2020, p. 53). Building on the BTM, the Business Model Canvas Model (BMCM) provides mechanisms to synchronise the transactional model into classical BMC representation. Overall, this tooling environment supports automatic generation of business plans. This research builds on this stream by proposing a comprehensive BMC metamodel grounded in UML and MOF, complemented by explicit constraints expressed in OCL and AQL, and supported by a dedicated DSML tool.

## 2.2. Metamodels from an Information System Perspective

While a *model* is a simplified representation of reality or an abstraction from reality (Teece, 2007, p. 1320), a *metamodel* defines instead a set of rules that specify concepts and govern the relationships between components used to create models (Jeusfeld, 2009). In other words, it is a model that allows the creation of models. Metamodelling is a long-standing and well-established practice within the IS community (Iivari & Koskela, 1987). Metamodels can be considered as language specifications in the Domain-Specific Modelling Language (DSML) field or as software specification in enterprise architecture tooling field (Fill et al., 2013). Conceptual modelling plays a central role in IS research (Lukyanenko et al., 2019). Metamodelling is a common method for conceptualisation (Lin et al., 2007, p. 351) and has a broad range of applications such as engineering, architecture, or 3D modelling (Fill et al., 2013; Kyriakou et al., 2017). A metamodel plays a critical role in modelling activities, as any modification can affect the validity of models based on it (Dietz & Juhrisch, 2012, p. 249). A metamodel can be used to describe ontologies. However, a metamodel differs from an ontology. An ontology is used to describe concepts (and instances) and their relationships. An ontology is primarily used in knowledge creation contexts. Furthermore, an ontology is a descriptive model (Aßmann et al., 2006). A concept is defined as set of ideas that are linked with logical rules (Hassan et al., 2019, p. 201). Concepts are defined through class definitions (Eriksson et al., 2019, p. 6). A relationship is a cause-and-effect link between concepts.

Graphical representations (e.g. UML) are well-suited for metamodel design, as they are simple to understand for both specialists and non-specialists (Dietz & Juhrisch, 2012, p. 231). Graphical representations such as those used in enterprise architecture help reduce technological complexity (Beese et al., 2023). The Unified Modelling Language (UML), a General-Purpose Modelling Language (GPML) (Frank, 2014), has a wide acceptance among practitioners (Dietz & Juhrisch, 2012, p. 246) and scholars (Allen & March, 2012; Lin et al., 2007; Lukyanenko et al., 2019) especially for enterprise modelling (Salem et al., 2008). Furthermore, UML is a flexible standard (Fernández-Medina et al., 2007) that improves appropriation by other researchers. The language enables the creation of different kinds of diagrams (e.g. class diagrams, activity diagrams). A *class* represents a set of objects sharing the same attributes, operations, methods, relationships, and semantics (Object Management Group, 2017, p. 194; Rittgen, 2006, p. 75). Class diagrams are well-suited to representing concepts (with their properties) and their relationships. The principal relationships are association (represents a link between instances of classes), generalisation (indicates that a class inherits from a parent class), composition (implies that an instance of a class (the composite) consists of instances (the components) that are also destroyed if the composite is destroyed), and aggregation (means that an instance (the composite) consists of instances (the components) that continue to exist even if the composite is deleted) (Object Management Group, 2017). When representing an association between classes, each side has a multiplicity that represents the potential number of values at that end when the value of one end is fixed (Allen & March, 2012, p. 954).

## 2.3. Metamodel Theoretical Foundations: from the $V^4$ Framework to the BMC Metamodel

This article builds on the $V^4$ Framework, which consists of four dimensions: value proposition, value architecture, value network, and value finance (Al-Debei & Avison, 2010). The Value Proposition dimension describes the value elements incorporated in the offering and their target market. The Value Architecture depicts how the company creates value in terms of resources and resource configuration. This dimension is deeply rooted in the RBV of the firm. The Value Network represents the relationships through which the company is involved. And finally, the Value Finance dimension describes financial performance elements (e.g. cost, pricing, revenue).

To improve readability and operationalisation of the framework, this research suggests three adaptations. First, value architecture and value network are merged into one dimension since relationships are considered relational resources in the RBV (Srivastava & Gnyawali, 2011). This change enhances the alignment of the framework with the RBV. It is in line with BMC authors that group together *KP*, *KR* and *KA* into architecture perspective (Fritscher & Pigneur, 2010). Moreover, a partner may allow access to external resources (Osterwalder et al., 2015, p. XVI). Secondly, the names of the components have been adapted to enhance readability and differentiation between the dimensions. Table 4 summarises how $V^4$ Framework has been adapted to design the BMC Metamodel.

| $V^4$ Framework | | BMC Metamodel | BMC components |
|---|---|---|---|
| Value Architecture Value Network | → | Key Element | Key Resources, Key Activities, Key Partnerships |
| Value Proposition | → | Value Element | Value Proposition, Customer Segment, Customer Relationship, Channels |
| Value Finance | → | Performance Element | Cost Structure, Revenue Streams |

*Table 4: $V^4$ Framework Adaptation*

The theoretical framework (Figure 1) of the BMC metamodel is based on a tripod: *Key Element* which represents the value source; *Value Element* which represents value creation; and Performance Element which represents the performance of the firm.

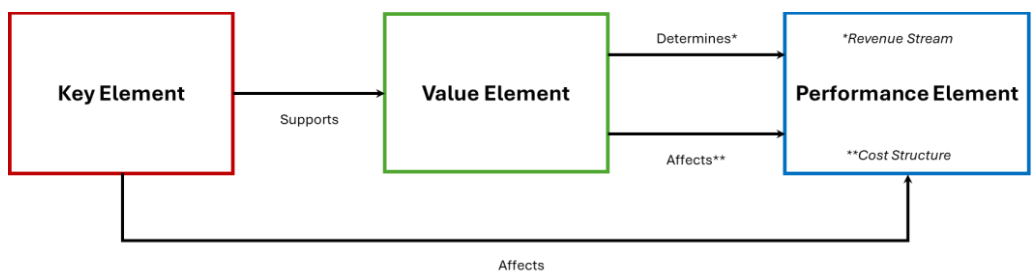

Thirdly, the relationships between the components have been defined precisely to enhance operationalisation. This operationalisation responds to prior calls for "*an engineering BM ontology in terms of elements, properties (relationships), constraints and rules, semantics and possibly notations might, for instance, reinforce understanding, while also facilitate the development of computer-based modelling tools that would be potentially helpful to practice*" (Al-Debei & Avison, 2010, p. 373).

## 3. Research Methodology

## 3.1. Research design

The main purpose of this research is the design and implementation of a Domain-Specific Modelling Language (DSML) for the Business Model Canvas (BMC). A modelling language consists of three core dimensions: syntax, semantics, and notation. The syntax specifies elements and attributes, and is commonly divided into abstract syntax and concrete syntax. The semantics assigns meaning to the concepts. The notation specifies the visual representation of the language (Fill et al., 2013).

In this research, the abstract syntax of the BMC DSML is defined using UML and Ecore class diagrams. Ontology-driven approaches such as OntoUML are not adopted, as the objective of this work is not ontological analysis but the formal specification of a DSML that can be directly operationalised through modelling tools. The semantics are grounded in existing knowledge of the BMC and of the relationships between its building blocks, and are formalised using UML and Ecore edge relationships, complemented by OCL (UML) and AQL (Ecore). Finally, the concrete syntax and visual notation are implemented using a Sirius-based graphical modelling environment, enabling the instantiation and validation of concrete business models.

Design science aims at solving real-world problems by applying existing knowledge (from behavioural sciences) from a practical perspective (Baskerville, 2008, p. 442; Hevner et al., 2004, p. 76). This research follows Design Science Research (DSR) principles to produce innovative Information Technology (IT) artefacts. Such IT artefacts may be: constructs (vocabulary and symbols), models (abstractions and representations), methods (algorithms and practices), and instantiations (implemented and prototype systems) (Hevner et al., 2004, p. 77). The process of DSR should be structured in five steps: 1) identifying a problem, 2) defining the research objective, 3) building the requirements, 4) developing and testing a solution, and 5) evaluating the solution (Dietz & Juhrisch, 2012, pp. 230-231).

Table 5 depicts each research step with its corresponding elements.

| DSR steps | Elements |
|---|---|

| Problem identification | - Most practitioners use the BMC as a checklist without paying attention to relationships among building blocks (Avdiji et al., 2020, p. 708).<br>- BMC is map-based instead of being network-based which limits the representation of explicit relationships (Szopinski et al., 2022, p. 785). |
|---|---|
| Research objective definition | - Provide a complete metamodel for BMC (abstract syntax) from MOF M2 level.<br>- Suggest a DSML tool for designing BMC (concrete syntax) from MOF M1 level.<br>- Provide a real-world example from construction industry using the IT artefact tool (instances) from MOF M0 level. |
| Requirement building | - Identify exhaustive relationships between BMC components.<br>- Easy to use BMC modelling tool.<br>- Based on open source software (Papyrus Designer, Ecore, Obeo Designer, and Sirius).<br>- Represent business model portfolio of companies. |
| Solution development and testing | - More than seven versions of the BMC Metamodel were developed and tested.<br>- Three major versions of the Ecore Metamodel were created and tested.<br>- BMC Modeler has been tested in more than 15 versions. |
| Solution evaluation | - Evaluation by one Enterprise Architecture (EA) scholar, by a CEO of a software-based company, and by an entrepreneur.<br>- Comparative evaluation of three types of BMC (BMC without relationship, BMC with arrows, and BMC Modeler created) using five AI (Mistral AI, Google Gemini, ChatGPT (with and without account), Perplexity Pro).<br>- Evaluation regarding BMC Modeler performance to represent a real-world business model. |

*Table 5: Process of the research*

Specifically, this research consists of the following steps: 1) reviewing the relevant literature on Business Model Canvas to identify classes and relationships; 2) grouping relationships in a spreadsheet based on meaning (at least seven iterations were necessary to find a satisfying solution); 3) translating concepts and relationships into UML; 4) creating the *Ecore* Model; 5) creating a Domain-Specific Modelling Language (DSML) tool. UML modelling was performed using *Papyrus Designer* (version 7.1.0) while Ecore and DSML modelling was performed using *Obeo Designer* (version 11.9). Papyrus Designer is a well-known tool from the Eclipse community that is based on the UML 2.5 specification. Obeo Designer is an open source software that integrates Sirius. Sirius technology allows the creation of domain-specific graphical modelling workbenches based on the Eclipse Modelling Framework (EMF) and the Graphical Modelling Framework (GMF).

The metamodel suggested in this research was designed using UML. This choice is based on good practices used in both industrial and academic settings. For instance, specifications BPMN 2.0 (Object Management Group, 2011, p. 55) and ArchiMate 3.2 (The Open Group, 2023, p. 7) describe their metamodels using UML or UML-like class diagram. UML 2.5.1 specification itself defines stereotype metaclass extensions using UML (Object Management Group, 2017, p. 261). Furthermore, the Meta Object Facility (MOF) uses the UML "*to describe modelling languages as inter-related objects*" (Object Management Group, 2014, p. 14). From the academic side, the e3value concepts and relationships are expressed using UML class diagrams (Akkermans & Gordijn, 2003, p. 119).

This research follows the three cycles of Design Science Research (Hevner et al., 2004). Table 6 depicts the DSR iterative process across the relevance, rigour, and design cycles (Ralyté et al., 2025). Dates are only indicative and reflect major design decisions.

| Date | DSR Cycle | Description |
|---|---|---|
| January 2025 | Relevance cycle | Initial discussion with an entrepreneurship professor addressing the limitations of modelling relationships between Business Model Canvas (BMC) components. |
| January 2025 | Design cycle | Design of a course on BMC structure (Element and Relationship). Artefacts: relationship table derived from BMC literature; first UML class diagram. |
| January 2025 | Rigour cycle | Evaluation of the initial artefact by an Enterprise Architecture (EA) professor. |
| February 2025 | Rigour cycle | Workshop on UML modelling with an EA professor to strengthen formal modelling foundations. |
| February 2025 | Design cycle | Design of the first version of the BMC metamodel for an interdisciplinary course combining MIS and Engineering Sciences. |
| February 2025 | Rigour cycle | UML modelling workshop (advanced concepts). |
| February 2025 | Relevance cycle | Request from former students (2015 cohort) creating a data-driven AI startup for support on business model design. |
| March 2025 | Relevance cycle | Receipt of the startup's first BMC version using a classical canvas template. |
| March - April 2025 | Relevance cycle | Iterative exchanges and online meetings with the three founders of the data startup. |
| February - June 2025 | Design cycle | Writing and submission of a first research article on BMC modelling (submitted to BISE on 5 June 2025). |
| July 2025 | Rigour cycle | Artefact evaluation through BISE peer review (reject with encouragement to resubmit). |
| August - November 2025 | Design cycle | Redesign of the BMC metamodel. Artefact: revised UML metamodel. |
| August - November 2025 | Rigour cycle | Methodological refinement based on reviewer feedback and modelling theory. |
| August - November 2025 | Design cycle | Implementation of the metamodel in Ecore and Sirius tooling. Artefact: first version of the BMC Modeler. |
| September 2025 | Rigour cycle | Evaluation of the redesigned artefact by an EA professor. |
| October 2025 | Relevance cycle | Modelling of real-world BMCs (Plastic Material Company, Data Startup, software-based company). |
| November 2025 | Design cycle | Metamodel refinement based on implementation experience and OCL constraints. |
| December 2025 | Rigour cycle | Workshop on MOF and OCL. Evaluation of the metamodel and OCL constraints. |
| December 2025 | Design cycle | Final improvement of the metamodel and OCL artefacts. |

*Table 6: Research trajectory across the Design Science Research cycles*

## 3.2. Metamodel Methodology

The proposed Business Model Canvas metamodel is grounded in the UML specification (Object Management Group, 2017). Its structure adopts the canonical directed-graph modelling pattern defined in UML specification, in which node and edge classes are connected through explicit source and target references (Object Management Group, 2017, p. 374). The metamodel was developed using *Eclipse Papyrus* (version 7.10), an open source modelling environment implementing UML version 2.5. The resulting class diagram is presented in Figure 2.

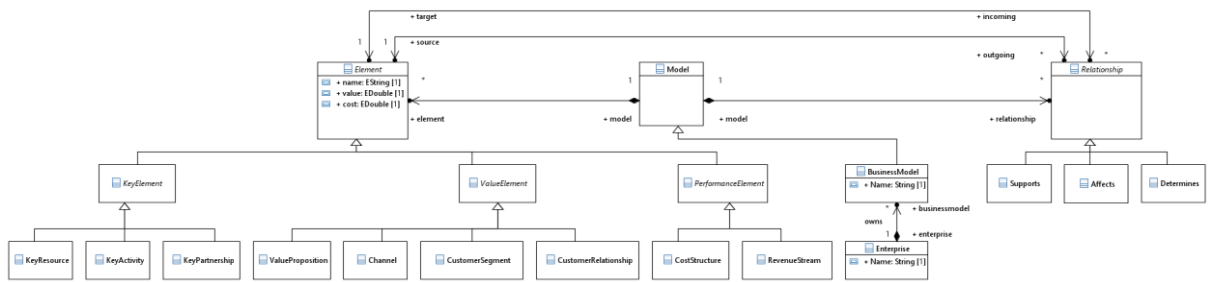

This metamodel is structured as follows. A model consists of an *Element* and a *Relationship* (*abstract classes*). A *BusinessModel* is a class that is a kind of *Model*. This means that, through inheritance a *BusinessModel* consists of *Element* and *Relationship*. An *Enterprise owns* one or more *BusinessModel*. *Element* class specialises into *KeyElement, ValueElement, and PerformanceElement.* KeyElement specialises as *KeyResource*, *KeyActivity*, and *KeyPartnership*. *ValueElement* specialises into *ValueProposition*, *Channel*, *CustomerSegment*, and *CustomerRelationship*. *PerformanceElement* specialises into *CostStructure* and *RevenueStream*. The *Relationship* class is divided into three relationship types[3] *Supports*, *Determines* and *Affects*. *Supports* means that the component provides a service to another. *Determines* means there is direct causality between components. *Affects* means there is indirect or complex causality between components. These three relationships were identified through three steps detailed in the next section. This metamodel is simple to enhance maintainability. Hence, all constraints are expressed using the Object Constraint Language (OCL).

*Matrix of component relationships*

Since the BMC includes nine components, all possible relationships can be represented in a 9$x$9 matrix. This results in 81 potential relationships. The self-relationships are considered as *supports*. As relationships are considered symmetric, only active-voice relationships need to be specified.

| | KR | KA | KP | CS | VP | CH | CR | R$ | C$ |
|---|---|---|---|---|---|---|---|---|---|
| **KR** | KR → KR | KR → KA | KR → KP | KR → CS | KR → VP | KR → CH | KR → CR | KR → R$ | KR → C$ |
| **KA** | KA → KR | KA → KA | KA → KP | KA → CS | KA → VP | KA → CH | KA → CR | KA → R$ | KA → C$ |
| **KP** | KP → KR | KP → KA | KP → KP | KP → CS | KP → VP | KP → CH | KP → CR | KP → R$ | KP → C$ |
| **CS** | CS → KR | CS → KA | CS → KP | CS → CS | CS → VP | CS → CH | CS → CR | CS → R$ | CS → C$ |
| **VP** | VP → KR | VP → KA | VP → KP | VP → CS | VP → VP | VP → CH | VP → CR | VP → R$ | VP → C$ |
| **CH** | CH → KR | CH → KA | CH → KP | CH → CS | CH → VP | CH → CH | CH → CR | CH → R$ | CH → C$ |
| **CR** | CR → KR | CR → KA | CR → KP | CR → CS | CR → VP | CR → CH | CR → CR | CR → R$ | CR → C$ |
| **R$** | R$ → KR | R$ → KA | R$ → KP | R$ → CS | R$ → VP | R$ → CH | R$ → CR | R$ → R$ | R$ → C$ |
| **C$** | C$ → KR | C$ → KA | C$ → KP | C$ → CS | C$ → VP | C$ → CH | C$ → CR | C$ → R$ | C$ → C$ |

*Table 7: BMC components relationships*

The first design rule (DR) for self-relationship is expressed in OCL.

```
-- DR 1: The relationship between an Element and itself must be Supports
context Relationship
inv Element_to_Self_is_Supports :
    (self.source = self.target)
    implies
    self.oclIsTypeOf(Supports)
```

---

[3] To simplify the diagram passive-voice forms are not included in the diagram (*is supported by*, *is determined by* and *is affected by*). Nevertheless, they represent the reverse relationship between each component.

*Relationship explicitation*

The relationships must be theory-grounded. Therefore, a literature review was conducted to clarify each relationship. Key references from BMC authors have been primarily used. Here some examples are given. The relationship between *Key Resource* and the *Value Proposition* (*KR → VP*) is clearly defined by the following citation: "*These resources allow an enterprise to create and offer a Value Proposition*" (Osterwalder & Pigneur, 2010, p. 34). The verb "*to allow*" implies "*to make possible to*", which corresponds to a **Supports** relationship. In other words, the *Key Resources* (*KR*) supports the *Value Proposition* (*VP*) in the sense that the *KR* makes it possible to the company to create and offer the *VP* to customers.

The relationship between *Key Activities* and the *Cost Structure* (*KA → C$*) is explicitly defined in: "*Creating and delivering value, maintaining Customer Relationships, and generating revenue all incur costs. Such costs can be calculated relatively easily after defining Key Resources, Key Activities, and Key Partnerships*" (Osterwalder & Pigneur, 2010, p. 40). Here the relationship between the *Costs Structure* (*C$*) and the *Key Activities* (*KA*) is rather an influence relationship. In other words, it means that the *KA* generates costs as part of value creation and delivery. This suggests an **Affects** relationship, as the impact of the *KA* on the *C$* is often indirect or difficult to quantify. Actually, one the *KA* may be linked to different products or services.

The relationship between *Value Proposition* and the *Customer Segment* (*VP → CS*) is defined here: "*Each Value Proposition consists of a selected bundle of products and/or services that caters to the requirements of a specific Customer Segment*" (Osterwalder & Pigneur, 2010, p. 22). Here the relationship between the *Value Proposition* (*VP*) and the *Customer Segment* is strong and implies a dependence. This implies **Determines** relationship.

*Relationships grouping*

To reduce complexity and enhance readability, the number of relationship types has been limited to three. The three relationships types (*Supports*, *Determines*, and *Affects*) have been identified through semantic analysis. Five iterations of the relationship table were required to achieve a meaningful and accurate classification. Table 8 explains the rules that have been applied for each type of relationship.

| Supports (is supported) | Determines (is determined by) | Affects (is affected by) |
| --- | --- | --- |
| ensures communication to | is created for | affects |
| is based on | generates | allows to earn |
| is delivered by | is determined by | contribute to |
| is ensured by | is generated by | influences |
| is sustained by | is generated from | is affected by |
| may be acquired from | addresses | is earned thanks to |
| may be ensured by | targets | is influenced by |
| may be the source of | | |
| may ensure | | |

| reaches | | |
|---|---|---|
| supports | | |
| sustains | | |
| performs | | |
| builds on | | |
| provides | | |
| enables | | |
| Is required by | | |
| reaches | | |
| established through | | |

*Table 8: classification of relationships*

In some cases, the relationship between two components was not clearly defined. Consequently, other sources have been used to complete relationships. Finally, only two relationships were supported by well-known real-world examples. These are *KP → CH* and *KP → CR*. *KP* may ensure final delivery (*CH*) (e.g. UPS and Amazon relationship). *KP* may ensure whole or part of *CR* for a company (e.g. Unisys and Dell relationship).

The *Element* class is further specialised into *Key Element* (that may be *Internal* or *External Key Element*), *Performance Element* and *Value Element*. This tripod is rooted in the RBV of the firm (Barney, 1991). The relationships between these components are summarised as follows. *Key Element (KE)* represents the value source. It supports value creation and delivery. KE can be considered the value means. *Value Element* (*VE*) determines *Revenue Streams* (*R$*) *Performance Element* (*PE*). Actually, *VE* depicts company intention that is to say, the potential value. When offered value is compared to competitors and assessed by potential clients it becomes the realised value throughout *Performance Element (PE)*. Besides, *VE* affects *Cost Structure* (*C$*) *PE* since to create value it is necessary to spend value. Furthermore, *KE* affects *PE* since resource is the main component that explains performance (Barney, 1991). Performance is the final component of the model since it represents the target of any business-oriented organisation (Al-Debei & Avison, 2010, p. 374; Baird & Raghu, 2015, p. 5). The internal *Key Element* represents resources, either internal (owned or controlled by the company) or external (mobilised from third parties), allowing the company to create and deliver value that determines its performance (Barney, 1991). The internal *Key Element* is either a *Key Resource* (*KR*) or a *Key Activity* (*KA*). The external *Key Element* consists of *Key Partnership* (*KP*).

*KE* consists of *Key Resources (KR)*, *Key Activities* (*KA*) and *Key Partnership* (*KP*). *KR* supports *KA* by providing essential services (Fritscher & Pigneur, 2010). *KR* can be considered critical from an operational perspective while *KA* is in turn resource-dependent. Sometimes, *KA* are supported by *KP* (Fritscher & Pigneur, 2010). This means that *KA* is partner-dependent, while partner is considered critical for operations. *KR* may also be owned by a *KP* (Fritscher & Pigneur, 2010), making the company dependent on its partner for access to resources, as a *KR* provider. Consequently, the following Design Rule (DR) is provided.

```
-- DR 2: The relationship between KE and KE must be Supports
context Relationship
inv KE_to_KE_is_Supports :
    (self.source.OclIsKindOf(KeyElement) and self.target.OclIsKindOf(KeyElement))
    implies
    self.OclIsTypeOf(Supports)
```

The *Value Element* is either a *Value Creation Element* or a *Value Delivery Element* (Osterwalder & Pigneur, 2010). The *Value Creation Element* is either *Customer Segment (CS)* or *Value Proposition (VP)*. *VP* is strongly linked to *CS*, as *VP* is typically created to address at least one *CS* (Osterwalder & Pigneur, 2010).

```
-- DR3: The relationship between CS and VP must be Determines
context Relationship
inv CustomerSegment_to_ValueProposition_is_Determines :
    (self.source.oclIsTypeOf(CustomerSegment) and
self.target.oclIsTypeOf(ValueProposition))
    implies
    self.oclIsTypeOf(Determines)
```

The *Value Delivery Element* is either a delivery *Channel (CH)* or *Customer Relationship (CR)*. *CH* ensures *CR* by enabling both *CS* communication and physical delivery.

```
-- DR4: The relationship between CH and VP must be Supports
context Relationship
inv Channel_to_ValueProposition_is_Supports :
    (self.source.oclIsTypeOf(Channel) and
self.target.oclIsTypeOf(ValueProposition))
    implies
    self.oclIsTypeOf(Supports)
```

The *Value Creation Element (VCE)* provides value to be delivered by the *Value Delivery Element (VDE)*. The component *Customer Segment (CS)* is the core of the BMC. *Value Proposition (VP)* is created for at least one *CS* (targeted CS). The company identifies and chooses which *CS* are targeted (Fritscher & Pigneur, 2010). In other words, *CS* determines *VP*. Besides, *VP* is affected by *Customer Relationship (CR)*. Strong *CR* increases perceived value, whereas weak *CR* may reduce it.

```
-- DR5: The relationship between CR and VP must be Affects
context Relationship
inv CustomerRelationship_to_ValueProposition_is_Affects :
    (self.source.oclIsTypeOf(CustomerRelationship) and
self.target.oclIsTypeOf(ValueProposition))
    implies
    self.oclIsTypeOf(Affects)
```

In addition, *VP* is delivered via *Channels (CH)* that represents the value delivery means (Fritscher & Pigneur, 2010; Osterwalder & Pigneur, 2010). *CH* also enables communication to *CS*, serving as customer touchpoints. *CS* is reached by *CH* (Fritscher & Pigneur, 2010). Indeed, *CS* expects a *CH* while the company offers at least one *CH*. Otherwise, *CH* supports *CS*.

```
-- DR6: The relationship between (CH or CR) and (CS or CR) must be Supports
context Relationship
inv
Channel_or_CustomerRelationship_to_CustomerSegment_CustomerRelationship_is_Support
s :
    (self.source.oclIsTypeOf(Channel) or
self.source.oclIsTypeOf(CustomerRelationship)) and
(self.target.oclIsTypeOf(CustomerSegment) or
self.target.oclIsTypeOf(CustomerRelationship))
    implies
    self.oclIsTypeOf(Supports)
```

The *Performance Element* is either a revenue (*Revenue Streams or R$*) or a Cost (*Cost Structure or C$*). Together, these elements allow performance to be assessed (Osterwalder & Pigneur, 2010). The *Performance Element* (PE) is either *Revenue Streams* (*R$*) or *Cost Structure* (*C$*). The configuration *R$* can affect *C$*. For instance, customers may pay using a particular means (e.g. PayPal) or a foreign currency. *C$* also affects revenues (*R$*). Generating *R$* requires incurring cost (*C$*). However, the exact contribution of cost to revenue generation remains difficult to assess.

```
-- DR7: The relationship between PE and PE must be Affects
context Relationship
inv PerformanceElement_to_PerformanceElement_is_Affects :
    (self.source.oclIsKindOf(PerformanceElement) and
self.target.oclIsKindOf(PerformanceElement))
    implies
    self.oclIsTypeOf(Affects)
```

*BMC Components relationships*

The relationships between each BMC components have also been specified using OCL. Nine perspectives are suggested. For each perspective, relationships within *KE* (between *KR*, *KA,* and *KP*), *VE* (between *CS*, *CR*, *VP*, *CH*) and *PE* (between *C$* and *R$*) are not described since they have already been specified in the previous section.

*Key Elements (KE) perspective*

As mentioned earlier, *KE* consists of *KR*, *KA* and *KP*. *KR* supports *VE* (*CS*, *CR*, *VP*, *CH*) and affects *PE* (*C$*, *R$*). Actually, *KR* supports *VP* since it provides services that distinguish firm's *VP* from its competitors. *KR* may also support *CH* because it provides for instance logistical or information services. *KR* supports *CS* and *CR*. *KR* may be technological or even relational resources that are critical to support customers needs or ensuring relationship.

*KR* affects costs(*C\$*) and revenues (*R\$*). Maintaining distinctive resources can be costly (Osterwalder et al., 2020, p. 164). *KR* contributes to generating revenues but the clear relationship between resources and financial performance is complex.

*KA* represents the operational side of the company. *KA* are supported by *Key Resources* (*KR*) (Fritscher & Pigneur, 2010). Indeed, *KA* are based on the assets and capabilities of the firm. *KR* is considered critical from an operational perspective while *KA* is in turn resource-dependent. Most of the time, firms are unable to perform all activities (*KA*) alone, instead, they rely on partners (*KP*) (Osterwalder & Pigneur, 2010, p. 39). In that situation, it means that *KA* is partner-dependent while partner is considered critical for operations. *KA* maintains relationships with *CS*. Furthermore, firm activities (*KA*) are essential for value creation (*VP)*, to value delivery (*CH*) and maintaining customer relationship (*CR*) (Osterwalder & Pigneur, 2010, p. 36). *KA* influences both revenue (*R\$*) and costs (*C\$*). Some activities (*KA*) may be critical to earn revenue (*R\$*). *KA* generates cost volume which affects *C\$* (Osterwalder & Pigneur, 2010, p. 40).

Most firms cannot fulfil their operations without external partners. *KP* describes required partnerships. *KP* supports *VP*, as it provides raw materials or services necessary to fulfil *VP* (Fritscher & Pigneur, 2010). Likewise, *KP* may provide services such as transportation to one or more *CH*. In some cases, *CR* may be wholly or partly outsourced to an external partner (*KP*). For instance, a computer manufacturer may outsource on-site repairs to a *KP*. Partnerships are at the center of value creation, as the firm defines which part of the value is given to the partner. Hence, *KP* affects revenues (*R\$*). Similarly, external services or the acquisition of raw materials affect costs (*C\$*).

```
-- DR8: The relationship between KE and VE must be Supports
context Relationship
inv KeyElement_to_ValueElement_is_Supports :
    (self.source.oclIsKindOf(KeyElement) and
self.target.oclIsKindOf(ValueElement))
    implies
    self.oclIsTypeOf(Supports)

-- DR9: The relationship between KE and PE must be Affects
context Relationship
inv KeyElement_to_PerformanceElement_is_Affects :
    (self.source.oclIsKindOf(KeyElement) and
self.target.oclIsKindOf(PerformanceElement))
    implies
    self.oclIsTypeOf(Affects)
```

*Value Element (VE) perspective*

VE consists of *VP*, *CH*, *CS*, and *CR*. *CS* brings together customers with homogeneous needs, behaviours or attributes (Osterwalder & Pigneur, 2010, p. 20). *CS* is the focus of any enterprise. *CS* are supported by *resources* (*KR*) and *activities* (*KA*) that may differ for each *CS*. In turn, *KR* and *KA* are critical for supporting *CS* needs (Fritscher & Pigneur, 2010). Customer relationships (*CR*), delivery (*CH*), or even some activities may rely on partnerships (*KP*). In this case, *KP* is critical for supporting *CS*. Supporting *CS* is a source of cost volume through the use of resources, activities, and maintaining relationships. *CS* affects the cost structure (*C$*). In turn, *CS* generates revenues (*R$*). *CS* is profitable when revenues exceed costs.

*CR* is essential for creating trustworthy relationships with customers. *CR* is supported by resources (*KR*) that may be relational or technological. *CR* is also supported by *KA*, as it provides operational services, solves customers problems, and helps in understanding their needs. *CR* may be outsourced to a partner such as for on-site intervention. *CR* is costly and influences costing (*C$*). *CR* contributes to improving perceived value by the customer, which may impact revenues (*R$*).

The *Value proposition* (*VP*) describes the value offered by the company. The content of *VP* is determined by the targeted *CS*. *VP* is delivered through *channels* (*CH*). In other words, *CH* is the means of delivery for *VP*. *VP* is affected by *CR*. In fact, a good relationship improves the perceived value, while a poor relationship decreases it. *VP* is supported by resources (*KR*) that provides critical services. *VP* is based on *KA* that are critical from an operational perspective. Most of the time, *VP* is supported by an external partner (*KP*) that plays a critical role in the creation or delivery of *VP* (Fritscher & Pigneur, 2010). *VP* affects costs because it is based on *Key Elements* (*KE*). *VP* determines the revenue streams (*R$*), as *R$* are directly generated from *VP*.

The channels (*CH*) are the distribution means that allow the company to deliver value. *CH* delivers the *VP* and constitutes customer touch point that is essential to support *customer relationship* (*CR*) (Fritscher & Pigneur, 2010). *CH* are the physical means thanks to which customers (*CS*) are reached. It means that *CH* supports *CS*. *CH* may be supported by *KP* which provides for instance logistics services. *CH* is also dependent to resources (*KR*) and operations (*KA*). *CH* affects costs (*C$*) and determines the generated revenues (*R$*).

```
-- DR10: The relationship between VE and C$ must be Affects
context Relationship
inv ValueElement_to_CostStructure_is_Affects :
    (self.source.oclIsKindOf(ValueElement) and
self.target.oclIsTypeOf(CostStructure))
    implies
    self.oclIsTypeOf(Affects)

-- DR11: The relationship between VE and R$ must be Determines
context Relationship
inv ValueElement_to_RevenueStream_is_Determines :
    (self.source.oclIsKindOf(ValueElement) and
self.target.oclIsTypeOf(RevenueStream))
    implies
    self.oclIsTypeOf(Determines)
```

*Performance Element (PE) perspective*

*PE* consists of *C$* and *R$*. *C$* describes cost volume required to realise its business model. Each component of the BMC incurs costs (Osterwalder & Pigneur, 2010). This means that all components affect the cost volume of the company. However, assessing each cost can be a tricky problem. Maintaining resources (*KR*), ensuring partner relationships (*KP*), conducting operational activities (*KA*), defining (*VP*) and delivering (*CH*) value, understanding customer needs (*CS*) or even ensuring relationship to customers (*CR*), lead to higher costs.

*R$* represents the value clients agree to pay to acquire products and services. *R$* are generated through *KA*, *KR*, and *KP*. However, it is difficult to assess the contribution of each element. *R$* is determined by the content of *VP*. *R$* is depends on *CS* that are targeted and which relationship (*CR*) the company has with each *CS*. *CH* is the means through which value is delivered, meaning it contributes directly to *R$*.

The table below summarises the BMC relationships.

<p align="center"><em>Table 9: BMC relationships summarised</em></p>

| | | KR | KA | KP | CS | VP | CH | CR | R$ | C$ |
|---|---|---|---|---|---|---|---|---|---|---|
| KE | KR | supports | supports (11) | is supported by (6) | supports (6) | supports (6) | supports (11) | supports (6) | affects (6) | affects (8) |
| | KA | is supported by (11) | supports | Is supported by (9) | supports (7) | supports (7) | supports (7) | supports (7) | affects (7) | affects (8) |
| | KP | supports (6) | supports (9) | supports | supports (10) | supports (9) | supports (13) | supports (14) | affects (15) | affects (8) |
| VE | CS | is supported by (6) | is supported by (7) | is supported by (10) | supports | determines (1) | is supported by (2, 3) | is supported by (4) | determines (5) | affects (8) |
| | VP | is supported by (6) | is supported by (7) | is supported by (9) | is determined by (1) | supports | is supported by (3) | is affected by (12) | determines (5) | affects (8) |
| | CH | is supported by (11) | is supported by (7) | is supported by (13) | supports (2,3) | supports (3) | supports | supports (3) | determines (16) | affects (8) |
| | CR | is supported by (6) | is supported by (7) | is supported by (14) | supports (4) | affects (12) | is supported by (3) | supports | determines (5) | affects (8) |
| PE | R$ | is affected by (6) | is affected by (7) | is affected by (15) | is determined by (5) | is determined by (5) | is determined by (16) | is determined by (5) | supports | affects (8) |
| | C$ | is affected by (8) | is affected by (8) | is affected by (8) | is determined by (8) | is determined by (8) | is affected by (8) | is affected by (8) | affects (8) | supports |

(1) Osterwalder & Pigneur (2010 : 23), Fritscher & Pigneur (2010 : 32)

(2) Osterwalder & Pigneur (2010 : 20), Fritscher & Pigneur (2010 : 32)

(3) Osterwalder & Pigneur (2010 : 26), Fritscher & Pigneur (2010 : 32)

(4) Osterwalder & Pigneur (2010 : 28), Fritscher & Pigneur (2010 : 32)

(5) Osterwalder & Pigneur (2010 : 30), Fritscher & Pigneur (2010 : 32)

(6) Osterwalder & Pigneur (2010 : 34), Fritscher & Pigneur (2010 : 32)

(7) Osterwalder & Pigneur (2010 : 36), Fritscher & Pigneur (2010 : 32)

(8) Osterwalder & Pigneur (2010 : 40)

(9) Osterwalder & Pigneur (2010 : 39), Fritscher & Pigneur (2010 : 32)

(10) Osterwalder & Pigneur (2010 : 38)

(12) Osterwalder & Pigneur (2010 : 29), Fritscher & Pigneur (2010 : 32)

(11) Barney (1991), Fritscher & Pigneur (2010 : 32)

(13) KP may assure final CH final delivery (e.g. UPS and Amazon relationship).

(14) KP may assure whole or part of CR for a company (e.g. Unisys and Dell relationship).

(15) Osterwalder & Pigneur (2010 : 32)

(16) Osterwalder & Pigneur (2010 : 27), Fritscher & Pigneur (2010 : 32)

### 3.3. Designing a Domain-Specific Modelling Language Tool for Business Model Canvas

The BMC Domain-Specific Modelling Language (DSML) tool is grounded in Model-Driven Architecture (MDA) principles (Salem et al., 2008; Tang, 2009). MDA is itself rooted in the Meta Object Facility (MOF) specification, which defines a four-layered modelling architecture ranging from meta-metamodelling (M3) to instance modelling (M0) (Object Management Group, 2016).

In contrast to the classical UML usage, where modelling activities typically operate from the M1 level (modelling) to the M0 level (instance modelling) by relying on the UML metamodel, this research operates across the M2 level (definition of the BMC metamodel), the M1 level (DSML-based modelling and tooling), and the M0 level (instantiation of concrete BMC instances using the DSML tool).

While Multilevel Modelling (MLM) extends MOF capabilities by introducing a recursive architecture that allows an arbitrary number of classification levels, and facilitates the reuse and integration of DSML artefacts (Frank, 2014), such a level of expressiveness is not required for the scope of this research. Therefore, the standard four-layer MOF architecture is sufficient to design research artefacts and ensure conceptual coherence. Table 10 classifies research artefacts according to the MOF framework.

| Level | Description | Artefacts |
|-------|-------------|-----------|
| M3 | Meta-Metamodel | None |
| M2 | Metamodel | BMC Metamodel (UML and Ecore specified) |
| M1 | Model | Sirius Model, BMC Modeler |
| M0 | Object | BMC example created using the BMC Modeler |

*Table 10: Research artefacts classified under MOF framework*

The BMC Metamodel designed in UML has been ported to Ecore. The Ecore BMC Metamodel has been designed in a trial-and-error settings as recommended in design science research (Hevner et al., 2004, p. 99).

The first version of the BMC metamodel is a simple translation of the first UML version to Ecore. Since UML and Ecore are quite similar, the Ecore version is as the UML one. This version has been abandoned since it was too complex to implement with Sirius.

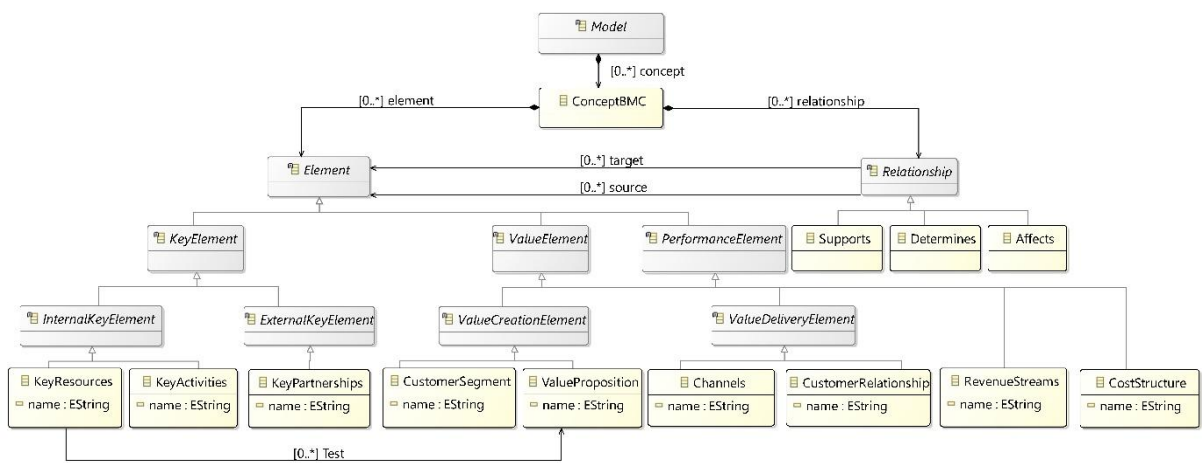

Conversely, the second version of the metamodel is simplistic to facilitate Sirius implementation. It consists of three classes: Element, Relationship, and a root class (ConceptBMC).

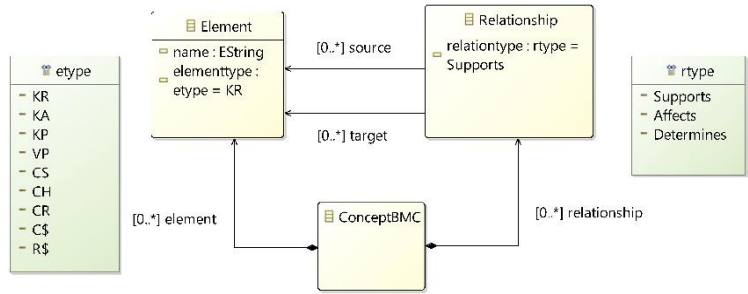

*Figure 4: BMC Ecore Metamodel version 2*

*Element* class is typed using enumeration to create BMC components. Similarly, *Relationship* has been typed using enumeration. *Element* and *Relationship* classes are linked using source and target relationship (UML association equivalent). This metamodel has been used to create a first version of the BMC Modeler. This version has been presented to an MDA expert and a BMC expert.

Based on the previous version feedback and testing, a third version of the metamodel has been designed to improve the BMC Modeler. It introduces two new classes (*Businessmodel* and *Enterprise*) and five relationships (composition). They are summarised below.

- *Element* consists of *Element*
- *Businessmodel* is composed of *Element*
- *Businessmodel* contains *Businessmodel*
- *Businessmodel* makes up *Relationship*
- *Enterprise* owns *Businessmodel*

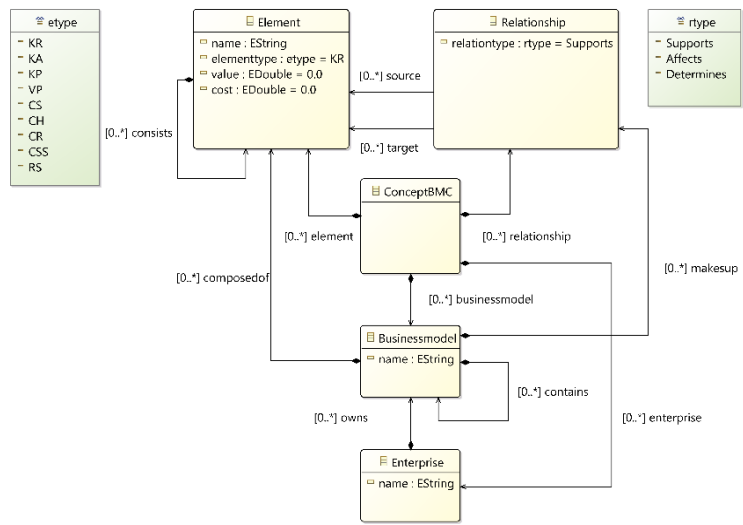

UML specified relationships has been implemented using The Acceleo Query Language (AQL). Rules have been implemented in a trial-and-error fashion. Four versions of AQL rules were necessary. For instance, the following rules have been implemented for *Determines* relationships.

**AQL rules for *Determines* relationship:**

*aql: (source.elementtype = bmc_emp::etype::CS and (target.elementtype = bmc_emp::etype::VP or target.elementtype = bmc_emp::etype::RS)) or ((source.elementtype = bmc_emp::etype::VP or source.elementtype = bmc_emp::etype::CH or source.elementtype = bmc_emp::etype::CR) and target.elementtype = bmc_emp::etype::RS)*

The BMC Modeler is an Eclipse application. It consists of a workspace and a palette. When the application is launched, it is mandatory to create a project. Then the user has to select the corresponding metamodel.

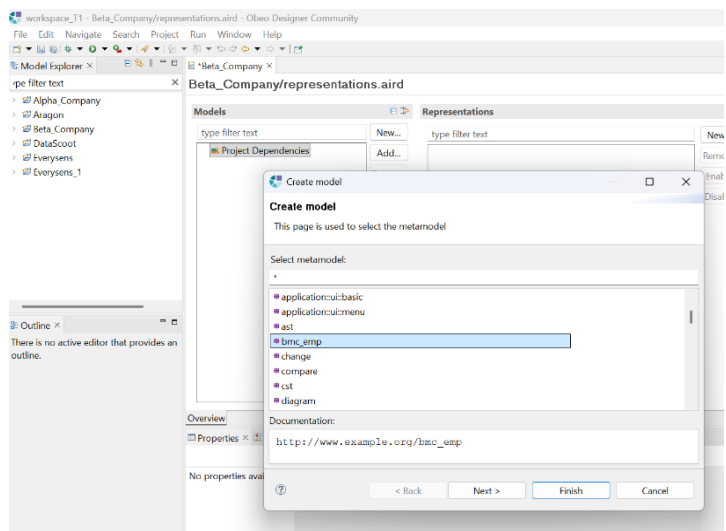

*Figure 6: Metamodel selection*

After that, the user has to create an enterprise (Figure 7).

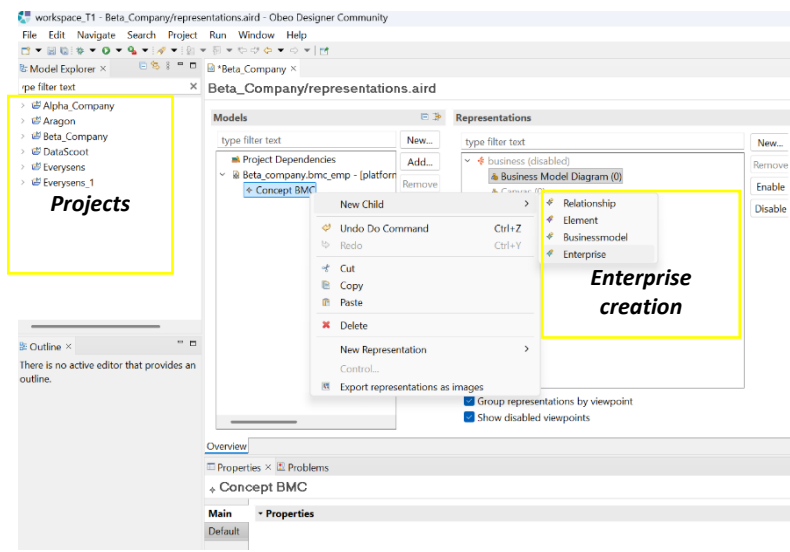

*Figure 7: Enterprise creation*

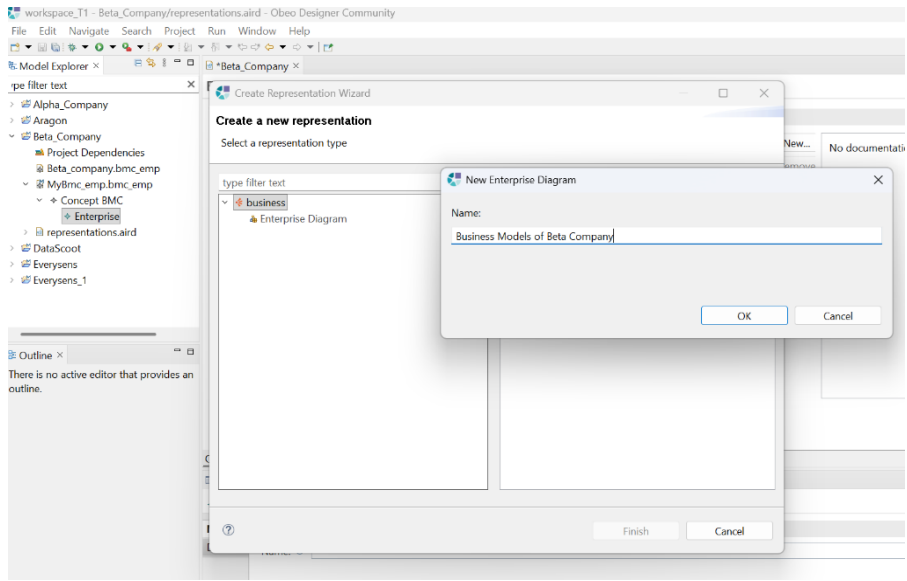

*Figure 8: Enterprise diagram creation*

Next the company is created. The user can create one or more business models. In the example, the Plastic Material Company (a real case study[4]) has four business models. To create a business model the user has to choose "*Create Enterprise Container*".

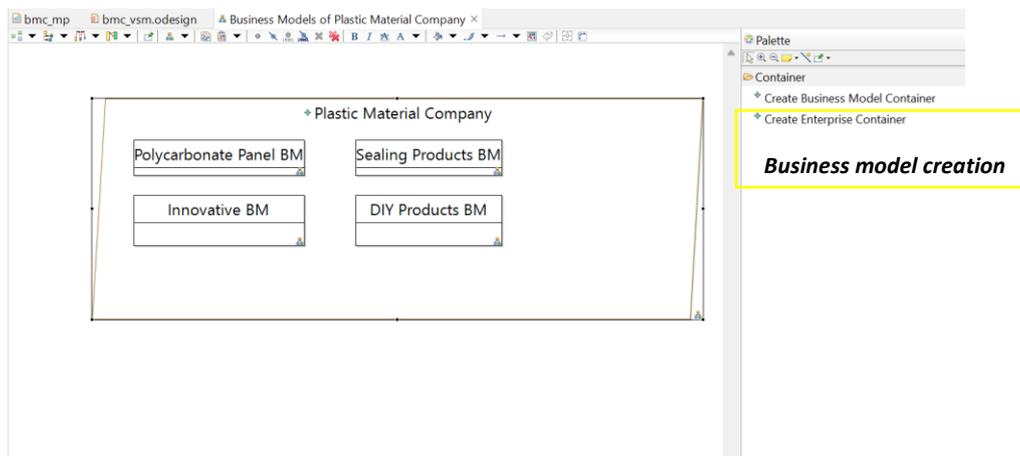

---

[4] The name of the company has been changed.

Following this the user can choose a business model to model it as wanted. The workspace (Figure 9) consists of the following components. On right had corner, there is the palette. It consists of the three-tier structure. Namely, *Key Element* (*KE*), *Value Element* (*VE*), and *Performance Element* (*PE*). Each category consists of corresponding BMC components. Each component is represented using an icon[5] (graphical representation) and the name of the component. Furthermore, relationships can be chosen to link the components. The AQL rules are applied to create or not the chosen relationship as in current modelling products (e.g. ArchiMate). For instance, it is possible to create a relationship *Supports* between *Key Activity* (*KA*) and *Value Proposition* (*VP*) while it is impossible to create a *Determines* relationship between the same elements (*KA* and *VP*).

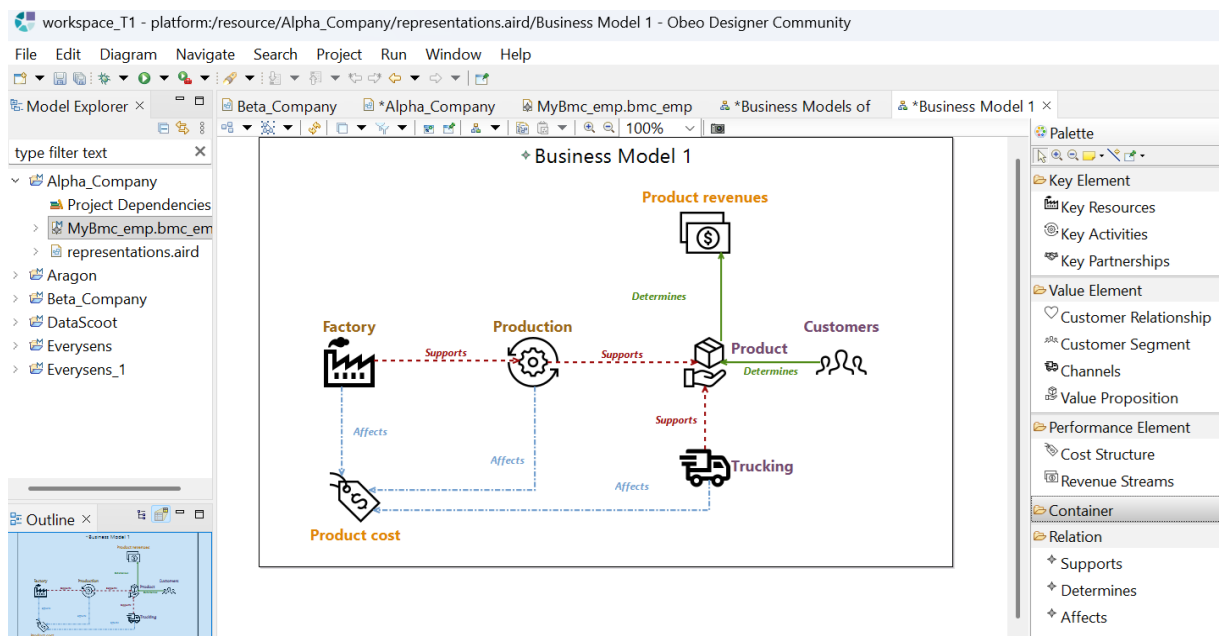

*Figure 9: Simplified example*

In this simplified example, the company has the following business model. A unique factory (*KR*) *supports* the production of goods (*KA*), which in turn *supports* product offering (*VP*). While customers need *determines* product characteristics (V*P*) and trucking activity (*CH*) *supports* product delivery. From a costing perspective, factory (*KR*), production (*KA*), and trucking (*CH*) *affects* product costs (*C$*). From a revenue perspective, product (*VP*) *determines* revenues (*R$*).

# 4. Assessing Research Artefacts Using Design Science Research Principles

## 4.1. Evaluation of the Business Model Canvas Modeler

BMC Modeler is the final artefact of this research, which has been assessed ex-post using both artificial and naturalistic evaluation. Artificial evaluation consists in assessing artefacts using artificially constructed case while naturalistic evaluation involves real case study, users, or problem (Ralyté et al., 2025). For the artificial evaluation, BMC Modeler diagram has been compared to original BMC with and without arrows. Furthermore, semantic analysis has been used to evaluate diagram richness and consistency. For the naturalistic evaluation, BMC Modeler has been confronted to a several case studies including the business model of a start-up in specialised in data visualisation and a company from construction industry.

---

[5] The free of rights icons have been downloaded from https://iconmonstr.com/

*Artificial evaluation: assessing BMC Modeler diagram using AI and semantic analysis*

To assess the BMC Modeler artefact over classic BMC representations on one hand, and to evaluate relationship pertinence, the following test has been conducted. The original template of the BMC has been used to represent the BMC produced in our example (see Figure 9). Two variants have been drawn: the first with neither relationships, nor arrows, and the second with arrows only. Then five Artificial Intelligence (AI) (ChatGPT, ChatGPT (without account), Perplexity Pro, Gemini Pro, and Mistral AI) have been asked to interpret the three BMC representations. The prompt used is as followed: "*Interpret this diagram for me, specifying in parentheses the abbreviations of the Business Model Canvas components that are related.*"

All AI produced paragraphs are available upon request. The results are the following. For all three representations, all AI identify BMC components. For the BMC without arrows, Nevertheless, all AI do not propose any relationship between components. They only describe what is inside each case of the BMC. They add some banal information about the BMC.

For the BMC with arrows AI behaviours are different. They describe relationships between components that are linked and describe some relationships between components that are not linked. To analyse text produced by AI, *Nvivo 15*, a qualitative data analysis software, has been used to classify the used verbs based on semantic analysis using the same methodology that has been used to identify the *Supports*, *Affects*, and *Determines* relationships. To normalise the analysis, passive voice has been converted into active voice forms. For instance, "*is created*", "create", "*creating*" were coded "*creates*" to reduce the number of nodes. To illustrate the process of coding a couple of examples are given here. Example 1: "*On the revenue side, the Product (VP) generates Product revenues (R$)*" (ChatGPT text) was coded as "*generates*" (between *VP* and *R$*). Example 2: "*production activities create a product that is delivered to customers through trucking*" (Perplexity Pro text) was coded as "*creates*" (between *KA* and *VP*) and "*delivers*" (between *CH* and *VP*). Example 3: A node regroups *Are* and *Represents*. It consists of the following sentence "*arrows from Factory, Production and Trucking show that resources, activities and channels are the main cost drivers of the business model (KR, KA, CH → C$)*" (Perplexity Pro). Similarly, "*The arrows point downward from both the Factory and Production to indicate that these resources and activities*

*represent the primary drivers of the Product cost found in the Cost Structure (C$)*" (Gemini 3 Pro). They are considered *Affects* relationships.

At the end of the process, 48 entries and 24 unique entries were coded and regrouped into four nodes: *Supports* for service meaning between component A to B; *Affects* for influence meaning between component A to B; *Determines* for a strong cause-consequence relationship and that implies a dependence between A to B; Other meaning for all other meaning. In fact, it was impossible to classify two nodes under *Supports*, *Affects*, and *Determines*. "*Purchases*" node consists of the following sentence: "*value proposition is created, effectively delivered through channels, and purchased by identified customer segments*" (Perplexity Pro). It describes customer behaviour not the business model. "*Captures*" node consists of the following sentence: "*Together, these relationships illustrate how the business creates value through internal operations on the left side of the canvas and captures value financially on the right side*" (Perplexity Pro). It describes a complex relationship that combines components (from *VP*, *CH*, *CS* to *R$*). The results are summarised in Table 11.

| Supports | Affects | Determines | Other meaning |
|---|---|---|---|
| Acts | Affects | Creates | Purchases |
| Communicates | Contribute | Defines | Represents |
| Delivers | Influences | Determines | |
| Enables | Are | Generates | |
| Meets | Captures | Shapes | |
| Provides | | | |
| Reaches | | | |
| Relies | | | |
| Serves | | | |
| Supports | | | |
| Transforms | | | |
| Transports | | | |

*Table 11: Semantic Coding of relationships for BMC with arrows*

The BMC Modeler artefact AI analysis reveals 61 codes and 21 unique entries. All the used verbs to describe component relationship has been classified into one of *Supports*, *Affects*, or *Determines* category. The results are summarised in Table 12.

| Supports | Affects | Determines |
|---|---|---|
| Delivers | Contributes | Defines |
| Distributes | Drives | Depends |
| Enables | Influences | Dictates |
| Ensures | Impacts | Generates |
| Offers | Links | |
| Provides | Shapes | |
| Reaches | | |

| Serves | | | |
|--------|--|--|--|

*Table 12: Semantic Coding of relationships for BMC Modeler*

The analysis of the BMC artefacts reveals several points. First, AIs are not able to identify any relationships between BMC components if they are not specified by arrows and/or text as shown in BMC without artefact analysis. Notably, this finding is consistent with previous research. Avidiji and colleagues (2020, p. 708) said that: "*A large number of practitioners used the Business Model Canvas as a checklist, using the nine building blocks as a list of aspects they needed to consider when designing a business model, without necessarily paying attention to the relationships between the solution elements and building blocks. For instance, practitioners might define a certain stream of revenue without relating it to a client segment.*" It is also true for scholars that use the BMC. Indeed, several researchers replicate the logic of BMC without following perspective knowledge (Avdiji et al., 2020, p. 708). Yet, BMC authors emphasise the role of relationships in BMC modelling (Avdiji et al., 2020, p. 707).

Secondly, the AI interpretation meaning when linking components (BMC with arrows only) is very close to the three main relationships that have been identified in theory building: *Supports*, *Affects*, and *Determines*. Furthermore, the more the relationships are specified the richer is the interpretative power of the diagram. Data shows that there are 27% more verbs used by AI to describe relationships with BMC Modeler Diagram (61 entries) than BMC with arrows (48 entries). Unique entries are with BMC Modeler Diagram (21 single entries on 61) and BMC with arrows (24 single entries on 48).

Considering a homogeneity ratio (HR) that can be calculated as follows.

$$Homogeneity\ ratio = \frac{Total\ entries}{Unique\ entries}$$

Results show *HR=2,9* for BMC Modeler and *HR=2* for BMC with arrows. Data show that the AI interpretation of the BMC Modeler diagram is 45% more homogeneous than the BMC with arrows diagram AI interpretation. Overall, these findings converge with prior research emphasising the impact of syntax on subjective and objective usefulness of business model representations (Szopinski et al., 2022, pp. 798-799).

*Natural evaluation: assessing BMC Modeler diagram using real-world case study*

To assess *BMC Modeler* performance, real-world business model has been modelled. The case study is based on Plastic Material Company[6], a small-sized European company (13 million euros turnover and 50 employees). PMC has been studied from May 2018 to December 2019 to create a multidisciplinary teaching case study. In contrast to action research, that is focused on problem-solving through organisational change, this research follows design science principles, that aiming at identify and solve real-world problems through design (Baskerville, 2008, p. 442). The dataset consists of observation, company's data (e.g. strategic vision of the CEO, sales data), and transcripts of seven in-depth interviews (two hours per interview). The dataset was under embargo until March 2023.

---

[6] The name of the company has been changed.

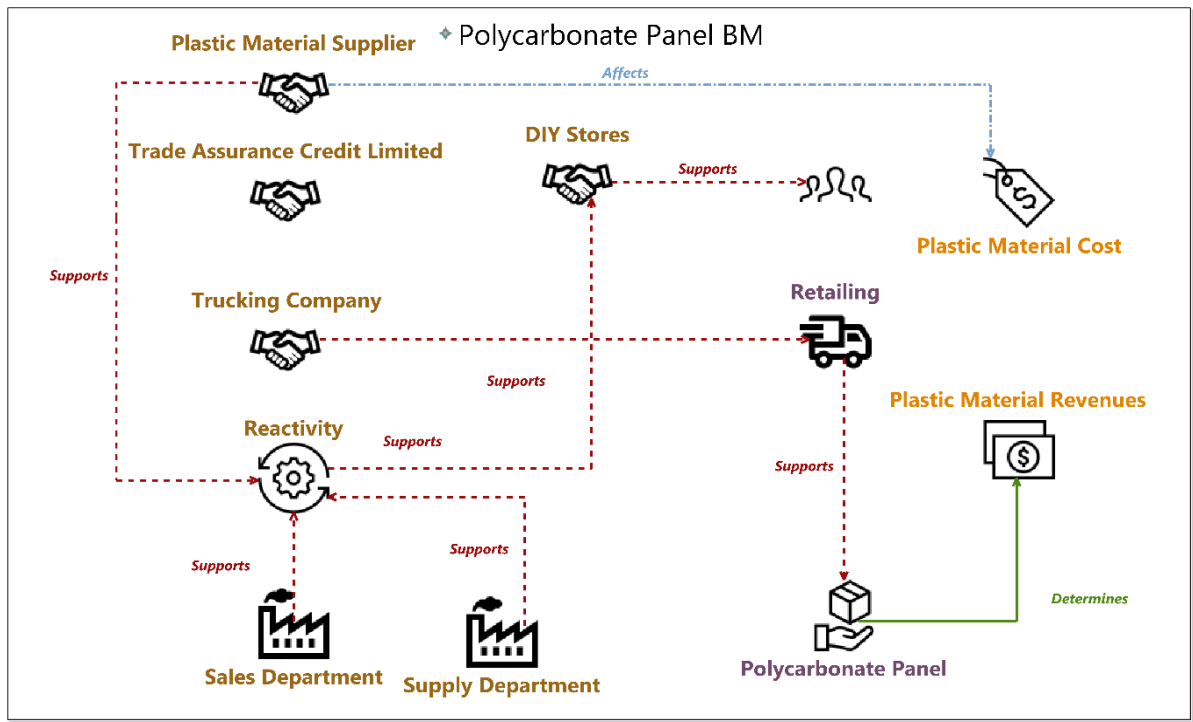

*Figure 10: PMC polycarbonate business model*

*PMC* business model is based on *Polycarbonate panels* (*VP*) that *PMC* sell to *DIY Stores* (*KP*). The margin is low for this kind of products because of a lack of differentiating elements. Consequently, competition is mainly based on price and reactivity. *Plastic Material Supplier* (*KP*) prices *affect* cost (*C$*) of *PMC*. *Reactivity* (*KA*) is necessary because exact sales prediction is almost impossible for *Sales Department* (*KR*). *Sales Department* (*KR*) and *Supply Department* (*KR*) both *supports Reactivity* (*KA*). Furthermore, *Plastic Material Supplier (KP) supports PMC reactivity* (*KA*). Consequently, *PMC* must maintain strong relationship with this *KP* to avoid shortages. While *PMC* credit insurance capacity relies on *Trade Credit Assurance Limited* (*KP*), it plays a secondary role in the business model. The *Supply Department* (*KR*) must prepare orders as quickly as possible. *DIY stores* (KP) *supports Private individual* (*CS*). In this business model, final customers (CS) are not in relationship with PMC.

## 4.2. Assessing the Research as Whole: A Design Perspective

There are three types of contributions in DSR: the design artefact, the foundations (knowledge base), and the methodology (Hevner et al., 2004, p. 87). This research contributes by providing design artefacts (BMC Metamodel, BMC Modeler tool), and by improving existing knowledge on the BMC (relationships).

This research provides the first fully specified metamodel for the BMC following Model Driven Architecture (MDA) methodology: metamodelling *M2* (BMC metamodel), modelling *M1* (BMC Modeler), and instancing *M0* (Plastic Material Company BMC). Furthermore, this research provides a clear causal-relationship structure for the BMC that operationalise the V$^4$ framework. It is a clear answer to previous call for engineering contribution to operationalises business models (Al-Debei & Avison, 2010; Veit et al., 2014). In fact, every business model is based on key structure (*Key Elements*) that allows value creation (*Value Element*) that explain firm performance (*Performance Element*). This tripod is rooted in Resource-Based View (RBV) of the firm and enhances theoretical foundation of BMC. Relationships between components is clearly necessary to fully understand a business model. Artificial evaluation (Ralyté et al., 2025), involving AI assessment, clearly proves that it is not sufficient to consider that BMC building block relationships are implicit. Indeed, previous research shows that practitioners uses the BMC as a checklist and do not pay any attention to relationships between components (Avdiji et al., 2020). Consequently, it is better to specify relationships especially to support strategic decisions. For instance, if a *Key Resource* (*KR*) is exploited by one business model, the abandon of such business model should imply resource disengagement or cession.

Secondly, this research provides an Eclipse plugin fully operational that allows rigourous diagramming. While designing the BMC using the official template can be useful in creative workshops (Avdiji et al., 2020), representing existing business models using only *map-based* models is harder. This research provides the *node-relationship* perspective that lack to the BMC (Szopinski et al., 2022). While business model modelling language (BMML) that are both map-based and node-relationship based is clearly scarce (Szopinski et al., 2022), this research contributes to existing knowledge on BMC.

Thirdly, thanks to the integration of design rules in BMC Modeler tool using Acceleo Query Language (AQL), it is possible to design canvases without knowing the specified rules. This is totally transparent for the end user while providing rich BMC representations. Furthermore, such node-relationship modelling obliges linking building blocks.

This research has been assessed using seven criteria provided by (Hevner et al., 2004).

| | Evaluation Criteria | Proof |
|---|---|---|
| | | |
| 1 | Problem Relevance | - The amount of research on Business Model is considerable: 16,911 articles of Springer Nature Link uses business model as keyword.<br>- The Business Model Canvas is considered a standard in both IS theory and practice (Avdiji et al., 2020; Lara Machado et al., 2024).<br>- A large part of practitioners uses the BMC without paying attention to relationships between components (Avdiji et al., 2020, p. 708). |
| 2 | Research Rigour | - A rigorous literature review has been conducted on Business Model formalisation and the Business Model Canvas.<br>- Research is based on $V^4$ framework (Al-Debei & Avison, 2010).<br>- Research is based on metamodelling (MOF, UML, OCL, Ecore) and graph-oriented modelling (MDA, GMF, AQL). |
| 3 | Design as a Search Process | - Relationships between BMC building blocks has been fully specified.<br>- Seven versions of the metamodel, three major versions of the Ecore metamodel, fourteen version of the BMC Modeler has been necessary to have a stable version of BMC Modeler. |
| 4 | Design as an Artefact | - While relationships between BMC building blocks are rather not fully specified, this research provides a clear specification of components relationships simplified through *Supports*, *Affects*, and *Determines* relationships.<br>- The performance of BMC Modeler over classic BMC template has been proved by Artificial Intelligence (AI) testing.<br>- The artefacts created are: a matrix with specified relationships, UML metamodel, OCL rules, Ecore Metamodel, BMC Modeler tool (Eclipse plugin). |
| 5 | Design Evaluation | - The first version of the artefacts has been assessed through: expert evaluation (one expert of enterprise architecture, a CEO of a software-based company with a PhD on enterprise architecture, a start-up founder).<br>- A comparative study of three kinds of BMC (BMC without arrow, BMC with arrows, and BMC Modeler created) using AI has been conducted.<br>- A real-world case study has been used to design a business model. |
| 6 | Research Contribution | - The first contribution is a fully specified BMC building blocks relationships that lacking in previous research (Avdiji et al., 2020).<br>- The second contribution is a fully operational Eclipse plugin that allows designing business model portfolios and BMC diagrams.<br>- The third is an implementation of prescriptive knowledge using OCL and AQL rules that are invisible for BMC Modeler users. |
| 7 | Research Communication | - The specified relationships improve existing knowledge on the Business Model Canvas.<br>- The advantage of fully specified relationships over without relationships have been adapted to managerial audience through managerial implications. |

## 5. Discussion

### 5.1. Positioning the Research Among Business Model Formalisation

Business models constitute a multidisciplinary topic spanning marketing, strategic management, and information systems (IS). Within the IS community, a wide range of conceptual and technical artefacts has been proposed to assist business model design and analysis (Lara Machado et al., 2024; Osterwalder & Pigneur, 2013). Among these artefacts, the Business Model Canvas (BMC) has emerged as a *de facto* standard for business models representation and communication (Avdiji et al., 2020).

Drawing on the $V^4$ framework (Al-Debei & Avison, 2010), this research adopts a structured theoretical perspective distinguishing *Key Element*, *Value Element*, and *Performance Element*. This framework explains firm performance through the interplay between internal and external resources, mediated by mechanisms of value creation. It therefore provides a coherent conceptual foundation for analysing and formalising the BMC within an IS perspective.

In this research, metamodelling is considered as a foundation for Design-Specific Modelling Language (DSML) rather than as a software specification activity (Fill et al., 2013). This research responds to prior call for engineering design approach to business models research (Al-Debei & Avison, 2010). As a modelling language, the DSML for BMC provides syntax, semantics, and notation (Fill et al., 2013). Table 13 summarises the theoretical contributions according to the criteria derived from Al-Debei and Avison.

| Expected contribution | Contribution |
|---|---|
| Elements | BMC components structured into Key Element (KE), Value Element (VE), and Performance Element (PE) superclasses |
| Properties (relationships) | Explicit BMC relationships (Supports, Affects, Determines) |
| Constrains and rules | Prescriptive knowledge operationalised through 11 design rules expressed in OCL and three sets of AQL rules integrated into a graphical modelling workbench |
| Semantics | Attributes associated to BMC components |
| Notation | Graphical representation of BMC components |

*Table 13: research contributions framed according to (Al-Debei & Avison, 2010)*

This research follows previous contributions aiming at formalise business model in general and BMC in particular. In contrast to prior approaches that either privilege visual simplicity (Avdiji et al., 2020), transactional precision (Curty & Fill, 2023; Razo-Zapata et al., 2015) or even business plan automation (Wieland & Fill, 2020), this research positions BMC formalisation as a Design Science Research (DSR) problem (Hevner et al., 2004) that requires explicit semantics while preserving usability for practitioners.

## 5.2. Benefits of the BMC Metamodel and BMC Modeler

The benefits of the proposed research artefacts are threefold. First, BMC metamodel contributes to existing knowledge on the operationalisation of the BMC (Fill, 2020; Lara Machado et al., 2024; Wieland & Fill, 2020). While BMC is widely adopted in practice, relationships between its components remain largely implicit and are often overlooked by both practicians and scholars (Avdiji et al., 2020). By explicitly defining relationships between BMC components, this research enhances the analytical expressiveness of the BMC and enables the development of dedicated tools as well as its potential integration into existing enterprise architecture tools such as ArchiMate (The Open Group, 2023).

Existing attempts to map the BMC to ArchiMate are fragmented and frequently contradictory as illustrated by the divergent interpretation of the *Key Activities* (*KA*) component (Fritscher & Pigneur, 2011; Iacob et al., 2014; Koyama et al., 2023; Walters, 2020). By providing a well-defined syntax, semantics, and notation for the BMC, the proposed metamodel establishes a necessary precondition for more consistent and rigorous mapping to other languages.

Thirdly, the BMC Modeler reconciles the ease of use traditionally associated to the BMC, which is essential for its practical relevance, with increased modelling rigour via the enforcement of formal rules using Acceleo Query Language (AQL). These rules prevent the user from creating incorrect relationships between components. As a result, users are not required to master the underlying metamodel explicitly to produce BMC instances that conform to the defined syntax and semantics. Overall, the combination of a formal metamodel and corresponding DSML tooling environment strengthens both the relevance and the rigour of BMC modelling, thereby addressing a critical gap in business model formalisation research.

### 5.3. Theoretical and Practical Implications

From a theoretical perspective, this research contributes to the business model formalisation literature by demonstrating that the explicit specification of relationships enhances the interpretative richness and consistency of business model representations. Conversely, business models relying only on implicit or weakly specified relationships exhibit reduced interpretative richness and consistency. As observed during the artificial evaluation (Ralyté et al., 2025), using artificial intelligence as an interpretative agent, models incorporating formally defined relationships led to more homogeneous interpretations than arrow-only models or classical BMC representations without explicit relationships. This suggests that the meaning of BMC diagrams is not solely conveyed by the special disposition of components, but also critically depends on the explicit semantics governing their interconnections.

From practical perspective, relationships between BMC components are often informally specified during BMC workshops, for instance through verbal explanations or the use of coloured Post-it notes (Avdiji et al., 2020). The proposed formalisation can therefore be understood as a mechanism for transforming such tacit relational knowledge into explicit, shareable modelling constructs.

This research highlights the central role explicit relationship specification in BMC use. BMC representations are commonly employed in entrepreneurship or intrapreneurship contexts, where clarifying business model mechanics is essential to convince investors and other stakeholders (Wieland & Fill, 2020). Prior research has shown that merely listing BMC components is insufficient to fully understand or communicate a business model (Avdiji et al., 2020; Wieland & Fill, 2020). By explicitly specifying relationships, the proposed DSML enhances readability and coherence of BMC representations. This approach aligns with prior research highlighting the impact of syntax on the subjective and objective usefulness of business model representations (Szopinski et al., 2022, pp. 798-799). Moreover, the integration of formal constraints, without requiring in-depth metamodelling expertise, improves modelling reliability while preserving the simplicity that characterises the BMC. Overall, this research demonstrates that rigour and relevance can be jointly achieved in business model modelling, without compromising usability.

### 5.4. Research Limitations and Future Research

This research indicates that explicitly specified relationships between BMC components lead to richer and more consistent interpretation than implicit or weakly specified relationships. One possible explanation for this finding can be drawn from knowledge creation theory (Nonaka, 1994), which distinguishes between tacit and explicit knowledge. While tacit knowledge is externalised through *externalisation* mechanisms (moving from tacit to explicit knowledge), explicit knowledge diffusion is rather based on *combination* mechanisms (moving from explicit-to-explicit knowledge).

However, this interpretation remains exploratory and calls for further empirical validation. In particular, controlled social experiments would be required to rigorously assess how different forms of relationship specification influence shared understanding, whether subjective or objective (Szopinski et al., 2022), among heterogeneous stakeholder groups.

A second limitation concerns the nature of the BMC Modeler itself. At this stage, the tool primarily constitutes a research artefact rather than a mature commercial application. Its implementation within Eclipse ecosystem may present usability challenges (e.g metamodel reference), particularly for users without prior experience in modelling environments. Moreover, although the DSML abstracts much of the underlying metamodel complexity, some conceptual understanding remains necessary to effectively use the tool.

Future research should therefore focus on the naturalistic evaluation (Ralyté et al., 2025) of the BMC Modeler with both novice and expert users. In particular, established technology adoption frameworks such as the Technology Acceptance Model (TAM) and its extensions (Davis, 1989; Venkatesh & Davis, 2000; Venkatesh et al., 2003), provide suitable instruments to assess perceived usefulness, ease of use, and behavioural intention.

Finally, concerning dissemination, a private GitHub repository has been established to support feature development and user feedback. As a next step, the BMC Modeler may be released as an open source project to foster broader adoption, transparency, community contribution (Von Krogh & Spaeth, 2007).

**Conclusion**

While prior research on business models and the Business Model Canvas (BMC) is abundant, several recurring criticisms have been raised in the IS and management research literature, such as a lack of concept clarity (Al-Debei & Avison, 2010), difficulties in representing business model dynamics (DaSilva & Trkman, 2014; Demil & Lecocq, 2010), or the absence of explicit relationships between components in BMC usage (Avdiji et al., 2020). Consequently, there remains substantial room for improvement in business model research (Al-Debei & Avison, 2010; Hedman & Kalling, 2003; Lara Machado et al., 2024).

The main objective of this article was to propose a metamodel for the BMC (Osterwalder & Pigneur, 2010). The proposed BMC metamodel is structured around three core building blocks: *Key Element* (*KE*), *Value Element* (VE), and *Performance Element* (*PE*). This theoretical framing is grounded in the V$^4$ Framework (Al-Debei & Avison, 2010) and rooted in the Resource-Based View (Barney, 1991). Relationships between these building blocks are formalised through a typology of relationships consisting of *supports* (foundational relationship), *determines* (direct causal link), and *affects* (indirect or systemic influences). The internal consistency of the BMC is further strengthened through explicit OCL constraints specifying admissible relationships between components.

This research leverages the complementary strengths of General-Purpose Modelling Language (GPML), namely UML and Ecore metamodelling, and Domain-Specific Modelling Language (DSML) (Frank, 2014) for end-user-oriented modelling through the BMC modeler. In doing so, it provides rigorous formal foundation that supports implementation and interoperability, and a usable modelling tool that deliberately hides unnecessary technical complexity for practitioners. Hence, this work demonstrates how rigour and relevance can be jointly achieved in business model modelling.

By clarifying the relationships between BMC components, this article opens new avenues for future research. A first direction concerns the empirical evaluation of the BMC Modeler with a broader population of users, including experts and novice practitioners. Moreover, this research contributes to the engineering of DSMLs tailored to business model modelling (Dietz & Juhrisch, 2012; Frank, 2014), and provides a structured foundation that may guide future IS research on BMC formalisation. Another promising research direction lies in the integration of the proposed metamodel into enterprise architecture frameworks such as ArchiMate, as suggested by prior work (Fritscher & Pigneur, 2010, 2011, 2015; Iacob et al., 2014). Given the growing adoption of ArchiMate (The Open Group, 2023) among practitioners and scholars (Lukyanenko et al., 2019), a rigorous language mapping between BMC and Archimate metamodels now becomes feasible. Such mapping integration would require model transformation techniques i.e., the process of converting one model into another (Fernández-Medina et al., 2007, p. 377).